\documentclass{jetpl}
\twocolumn
\lat

\usepackage[dvips]{graphicx}
\usepackage{amssymb,amsmath,epsf,slashed,cite}
\usepackage{wrapfig}

\def\pp{{\mbox{\boldmath$p$}}}
\def\thetap{\theta_{\mbox{\bfseries\itshape\scriptsize p}}}
\def\ep{E_{\mbox{\bfseries\itshape\scriptsize p}}}
\def\q{{\mbox{\boldmath$q$}}}

\def\cm{^{\rm \mbox{\tiny CM}}}

\def\p'{{\mbox{\boldmath$p'$}}}
\def\k{{\mbox{\boldmath$k$}}}
\def\ek{E_{\mbox{\bfseries\itshape\scriptsize k}}}
\def\l{{\mbox{\boldmath$l$}}}

\def\vz{{\mbox{\boldmath$0$}}}
\def\thetaps{\theta_{\mbox{\bfseries\itshape\scriptsize {p$^*$}}}}

\title{Final state interaction effects in electrodisintegration \\ of the deuteron within the Bethe-Salpeter approach}

\rtitle{Final state interaction effects\ldots}

\sodtitle{Final state interaction effects in electrodisintegration
of the deuteron within the Bethe-Salpeter approach}

\author{S.\,G.\,Bondarenko,
V.\,V.\,Burov, E.\,P.\,Rogochaya\/\thanks{e-mail: rogoch@theor.jinr.ru}}

\rauthor{S.\,G.\,Bondarenko S.G., V.\,V.\,Burov V.V., E.\,P.\,Rogochaya E.P.}

\sodauthor{Bondarenko, Burov, Rogochaya}

\address{Joint Institute for Nuclear Research,
141985 Dubna, Russia}

\dates{12 September 2011}{...}

\abstract{The electrodisintegration of the deuteron is considered
within a relativistic model of nucleon-nucleon interaction based
on the Bethe-Salpeter approach with a separable interaction
kernel. The exclusive cross section is calculated within the impulse approximation under various
kinematic conditions. Final state
interactions between the outgoing nucleons are taken into account.
The comparison of nonrelativistic and relativistic calculations is
presented. Partial-wave states of the neutron-proton pair with
total angular momentum $J=0,1$ are considered. }


\begin{document}

\maketitle

\section{Introduction}
The electrodisintegration of the deuteron is a useful instrument
which makes it possible to investigate the electromagnetic
structure of the neutron-proton ($np$) system. Many approaches
have been elaborated to describe this reaction for last 40 years
\cite{Forest:1983vc,Arenhovel:1982rx,Shebeko,Gakh:2004zq,Jeschonnek:2008zg}.
The simplest of them considered the electrodisintegration within a
nonrelativistic model of the nucleon-nucleon (NN) interaction and
outgoing nucleons were supposed to be free \cite{Forest:1983vc}
(the plane-wave approximation - PWA). Those approaches were in a
good agreement with experimental data at low energies. However,
further investigations have shown that the final state interaction
(FSI) between outgoing nucleons, two-body currents and other
effects should be taken into account to obtain a reasonable
agreement with existing experimental data at higher energies. Most
of these effects have been considered within nonrelativistic
models \cite{Arenhovel:1982rx,Shebeko}. In relativistic models, FSI effects could be calculated
within quasipotential approaches
including the on-mass-shell nucleon-nucleon $T$ matrix \cite{Gakh:2004zq,Jeschonnek:2008zg}.

One of fundamental approaches for a description of the $np$ system
is based on using the Bethe-Salpeter (BS) equation
\cite{Salpeter:1951sz}. To solve this equation we use a separable
ansatz \cite{Yamaguchi:1954mp} for the NN interaction kernel. In
this case, we have to deal with a system of algebraic equations
instead of integral ones \cite{Bondarenko:2002zz}. A separable
kernel model within the BS approach was not used to describe most
of high-energy NN processes for a while since calculated
expressions contained nonintegrable singularities. The separable
kernels proposed in
\cite{Bondarenko:2008fp,Bondarenko:2010qv,Bondarenko:2008mm} make
it possible to avoid those difficulties. Using these kernels FSI
can be taken into account in the electrodisintegration in a wide
range of energy. The electrodisintegration cross section is
calculated in the present paper under different kinematic
conditions (see Table \ref{tab:2} below). The rank-six NN
interaction potential MY6 \cite{Bondarenko:2010qv} is used to
describe scattered and bound $^3S_1$-$^3D_1$ partial-wave states.
The uncoupled partial-wave states with total angular momentum
$J=0,1$ ($^1S_0$, $^1P_1$, $^3P_0$, $^3P_1$) are described by
multirank separable potentials \cite{Bondarenko:2008mm}. The
obtained results are compared with nonrelativistic model
calculations \cite{Shebeko} where NN interactions were described
by the realistic Paris potential \cite{Lacombe:1980dr}.

The paper is organized as follows. In Sec.\ref{sec:2}\, , the
three-differential cross section of the $d(e,e^\prime p)n$
reaction is considered in the relativistic impulse approximation.
The used BS formalism is presented in Sec.\ref{sec:3}. The details
of calculations are considered in Sec.\ref{sec:4}. Then the
obtained relativistic results are compared with nonrelativistic
ones \cite{Shebeko} and experimental data of the Sacle experiment
\cite{Bussiere:1981mv,TurckChieze:1984fy} in Sec.\ref{sec:5}.
\section{Cross section}\label{sec:2}
When all particles are unpolarized the exclusive
electrodisintegration of the deuteron $d(e,e^\prime p)n$ can be
described by the differential cross section in the deuteron rest
frame - laboratory system (LS) - as:
\begin{eqnarray}
&&\frac{d^3\sigma}{dE_e'd\Omega_e'd\Omega_p}=
\frac{\sigma_\textmd{Mott}}{8M_d(2\pi)^3}{\frac{\pp_p^2{\sqrt s}}{\sqrt{1+\eta}|\pp_p|-E_p\sqrt{\eta}
\cos\theta_p}}
\nonumber\\
&&\times\left[ l^0_{00}W_{00}+l^0_{++}(W_{++}+W_{--})+ 2 l^0_{+-}\cos2\phi~{\rm Re}
W_{+-}\right.\nonumber\\
&&\hspace*{5mm}- 2 l^0_{+-}\sin2\phi~{\rm Im}
W_{+-}-2 l^0_{0+}\cos\phi~{\rm Re} (W_{0+}-W_{0-})\nonumber\\
&&\hspace*{5mm}\left.
-2 l^0_{0+}\sin\phi~{\rm Im} (W_{0+}+W_{0-})\right]. \label{3cross_0}
\end{eqnarray}
where $\sigma_{\rm
Mott}=(\alpha\cos\frac{\theta}{2}/2E_e\sin^2\frac{\theta}{2})^2$
is the Mott cross section, $\alpha={e^2}/{(4\pi)}$ is the fine
structure constant; $M_d$ is the mass of the deuteron;
$q=p_e-p_e'=(\omega,\q)$ is the momentum transfer; $p_e=(E_e,\l)$
and $p_e'=(E_e',\l')$ are initial and final electron momenta,
respectively; $\Omega_e'$ is the outgoing electron solid angle;
$\theta$ is the electron scattering angle. The outgoing proton is
described by momentum $\pp_p$ ($E_p=\sqrt{\pp_p^2+m^2}$, $m$ is
the mass of the nucleon) and solid angle
$\Omega_p=(\theta_p,\phi)$ where $\theta_p$ is the zenithal angle
between the $\pp_p$ and $\q$ momenta and $\phi$ is the azimuthal
angle between the ({\bf ee$^\prime$}) and ({\bf qp}) planes. Factor $\eta=\q^2/s$ can be calculated through the $np$ pair total
momentum $P$ squared:
\begin{eqnarray}
s=P^2=(p_p+p_n)^2
=M_d^2+2M_d\omega+q^2,
\label{pair_s}
\end{eqnarray}
defined by the sum of the proton $p_p$ and neutron $p_n$ momenta.
The photon density matrix elements have the following form:
\begin{eqnarray}
&&l_{00}^0=\frac{Q^2}{\q^2},\quad l_{0+}^0=\frac{Q}{|\q|\sqrt
2}\sqrt{\frac{Q^2}{\q^2}+\tan^2 \frac{\theta}{2}},\nonumber\\
&&l_{++}^0=\tan^2\frac{\theta}{2}+\frac{Q^2}{2\q^2},\quad
l_{+-}^0=-\frac{Q^2}{2\q^2},
\end{eqnarray}
where $Q^2=-q^2$ is introduced for convenience. The hadron density
matrix elements
\begin{eqnarray}
W_{\lambda\lambda'}=W_{\mu\nu}\varepsilon_\lambda^\mu\varepsilon_{\lambda'}^\nu,
\end{eqnarray}
where $\lambda$, $\lambda'$ are photon helicity components
\cite{Dmitrasinovic:1989bf}, can be calculated using the photon
polarization vectors $\varepsilon$ and Cartesian components of
hadron tensor
\begin{eqnarray}
W_{\mu\nu}=\frac{1}{3}\sum_{s_ds_ns_p}\left|<np:SM_S|j_\mu|d:1M>\right|^2,
\label{ht}
\end{eqnarray}
where $S$ is the spin of the $np$ pair and $M_S$ is its
projection; $s_d$, $s_n$ and $s_p$ are deuteron, neutron and
proton momentum projections, respectively. The matrix element
$<np:SM_S|j_\mu|d:1M>$ can be constructed according to the
Mandelstam technique \cite{Mandelstam:1955sd} and has the
following form in LS:
\begin{eqnarray}
&&<np:SM_S|j_\mu|d:1M>=i\sum_{n=1,2}\int\frac{d^4p\cm}{(2\pi)^4} \times\label{cur_fsi}\\
&&{\rm Sp}\left\{\Lambda({\cal L
}^{-1})\bar\psi_{SM_S}(p\cm,{p^*}\cm;P\cm) \Lambda({\cal L
})\Gamma_\mu^{(n)}(q)\right.\times \nonumber\\
&&S^{(n)}\left(\frac{K_{(0)}}{2}-(-1)^np-\frac q2\right)
\left.\Gamma^M
\left(p+(-1)^n\frac{q}{2};K_{(0)}\right)\right\}\nonumber
\end{eqnarray}
within the relativistic impulse approximation. The sum over
$n=1,2$ corresponds to the interaction of the virtual photon with
the proton and with the neutron in the deuteron, respectively. The
total $P\cm$ and relative ${p^*}\cm$ momenta of the outgoing
nucleons and the integration momentum $p\cm$ are considered in the
final $np$ pair rest frame - center-of-mass system (CM), $p$
denotes the relative $np$ pair momentum in LS. To perform the
integration, momenta $p$, $q$ and the deuteron total momentum
$K_{(0)}=(M_d,\vz)$ in LS are written in CM using the Lorenz-boost
transformation ${\cal L}$ along the $\q$ direction. The $np$ pair
wave function $\psi_{SM_S}$ is transformed from CM to LS by the
corresponding boost operator $\Lambda$. A detailed description of
$\psi_{SM_S}$, the $n$th nucleon interaction vertex
$\Gamma_\mu^{(n)}$, the propagator of the $n$th nucleon $S^{(n)}$,
and the deuteron vertex function $\Gamma^M$ can be found in our
previous works \cite{Bondarenko:2002zz,Bondarenko:2005rz}.

\section{Separable kernel of NN interaction}\label{sec:3}
The outgoing $np$ pair is described by the $T$ matrix which can be
found as a solution of the inhomogeneous Bethe-Salpeter equation
\cite{Salpeter:1951sz}:
\begin{eqnarray}
&&T(p^{\prime}, p; P) = V(p^{\prime}, p; P) \label{t_op} \\
&&\hspace*{10mm}+ \frac{i}{4\pi^3}\int d^4k\, V(p^{\prime}, k;
P)\, S_2(k; P)\, T(k, p; P),\nonumber
\end{eqnarray}
where $V$ is the NN interaction kernel, $S_2$ is the free
two-particle Green function:
\begin{eqnarray}
S_2^{-1}(k; P)=\bigl(\tfrac12\:{\slashed P}+{\slashed
k}-m\bigr)^{(1)} \bigl(\tfrac12\:{\slashed P}-{\slashed
k}-m\bigr)^{(2)},
\end{eqnarray}
and $p~(p')$ is the relative momentum of initial (final) nucleons,
$P$ is the total $np$ pair momentum.

To solve the BS equation (\ref{t_op}) partial-wave decomposition
\cite{Kubis:1972zp} for the $T$ matrix:
\begin{eqnarray}
T_{\alpha\beta,\gamma\delta}(p^{\prime},p; {P_{(0)}}) =
&&\sum_{JMab} t_{ab}(p_0',|\p'|;p_0,|\pp|; s) \times\label{bse_spd}\\
&&\hspace*{-7mm}({\cal Y}_{aM}(-{\p'})U_C)_{\alpha\beta}\otimes
(U_C {\cal Y}^{\dag}_{bM}({\pp}))_{\delta\gamma}\nonumber
\end{eqnarray}
is used. Here $P_{(0)}=(\sqrt s,\vz)$ is the $np$ pair total
momentum in CM, $U_C=i\gamma^2\gamma^0$ is the charge conjugation
matrix. Indices $a,b$ correspond to the set $^{2S+1}L_J^\rho$ of
spin $S$, orbital $L$ and total $J$ angular momenta, $\rho=+$
defines a positive-energy partial-wave state, $\rho=-$ corresponds
to a negative-energy one. Greek letters
$\{\alpha,\beta,\gamma,\delta\}$ in (\ref{bse_spd}) are used to denote Dirac matrix
indices. The spin-angle functions:
\begin{eqnarray}
&&{\cal Y}_{JM:LS {\rho}}(\pp) U_C= \label{an_pa}
\\
 && i^{L}\sum_{m_Lm_Sm_1m_2\rho_1\rho_2}C_{\frac12 \rho_1
\frac12 \rho_2}^{S_{\rho} {\rho}} C_{L m_L S m_S}^{JM} C_{\frac12
m_1 \frac12 m_2}^{Sm_S}\nonumber\\
&&\hspace*{20mm}\times Y_{L{m_L}}(\hat\pp)
{U^{\rho_1}_{m_1}}^{(1)}(\pp) {{U^{\rho_2}_{m_2}}^{(2)}}^{T}(-\pp)
\nonumber
\end{eqnarray}
are constructed from the free nucleon Dirac spinors $u,~v$. It
should be mentioned that only positive-energy states with $\rho=+$
are considered in this paper. Using a similar decomposition for
$V$, the BS equation for radial parts of the $T$ matrix and kernel $V$
is obtained:
\begin{eqnarray}
&&t_{ab}(p_0', |\p'|; p_0, |\pp|; s) = v_{ab}(p_0', |\p'|; p_0,
|\pp|; s) \label{BS_decomp}\\
&&+ \frac{i}{4\pi^3}\sum_{cd}\int\limits_{-\infty}^{+\infty}\!
dk_0\int\limits_0^\infty\! \k^2 d|\k|\, v_{ac}(p_0', |\p'|; k_0,
|\k|; s)\nonumber\\
&& \hspace*{20mm}\times S_{cd}(k_0,|\k|; s)\, t_{db}(k_0,|\k|;p_0,|\pp|;
s).\nonumber
\end{eqnarray}
To solve the resulting equation (\ref{BS_decomp}) a separable
ansatz \cite{Yamaguchi:1954mp} for the interaction kernel $V$ is
used:
\begin{eqnarray}
&&v_{ab}(p_0', |\p'|; p_0, |\pp|; s)=\label{ansatz}\\
&& \hspace*{10mm} \sum_{i,j=1}^N\lambda_{ij}(s) g_i^{[a]}(p_0',
|\p'|)g_j^{[b]}(p_0, |\pp|),\nonumber
\end{eqnarray}
where $N$ is a rank of a separable kernel, $g_i$ are model
functions, $\lambda_{ij}$ is a parameter matrix. Substituting $V$
(\ref{ansatz}) in BS equation (\ref{t_op}), we obtain the $T$
matrix in a similar separable form:
\begin{eqnarray}
&&t_{ab}(p_0', |\p'|; p_0, |\pp|; s)= \label{t_separ}\\
&& \hspace*{10mm}\sum_{i,j=1}^N\tau_{ij}(s)g_i^{[a]}(p_0', |\p'|)
g_j^{[b]}(p_0, |\pp|)\nonumber
\end{eqnarray}
where:
\begin{eqnarray}
\tau_{ij}(s)=1/(\lambda_{ij}^{-1}(s)+h_{ij}(s)),
\end{eqnarray}
and
\begin{eqnarray}
&&h_{ij}(s)= \label{hij}\\
&& -\frac{i}{4\pi^3}\sum_{a}\int dk_0\int \k^2d|\k|
\frac{g_i^{[a]}(k_0,|\k|)g_j^{[a]}(k_0,|\k|)}{(\sqrt
s/2-\ek+i\epsilon)^2-k_0^2}\nonumber
\end{eqnarray}
are auxiliary functions, $\ek=\sqrt{\k^2+m^2}$. Thus, the problem
of solving the initial integral BS equation (\ref{t_op}) turns out
to finding the $g_i$ functions and parameters $\lambda_{ij}$ of
separable representation (\ref{ansatz}). They can be obtained
from a description of observables in $np$ elastic scattering
\cite{Bondarenko:2010qv,Rupp:1989sg,Bondarenko:2008mm,Mathelitsch:1981mr,Bondarenko:2011hs}.

\section{Final state interaction}\label{sec:4}
In our previous works \cite{Bondarenko:2005rz,Bondarenko:2010qv}
the electrodisintegration of the deuteron was considered within
the plane wave approximation (PWA) when the final nucleons were
supposed to escape without interaction. Even though PWA is enough
to describe the electrodisintegration at low energies, it is important to take into account
the final state interaction (FSI) between the outgoing nucleons.
As it was shown in other works \cite{Shebeko}, the contribution of
FSI effects increases with increasing nucleon energies or/and
momentum transfer. Therefore, the relativistic models of the NN interaction must be elaborated and FSI should be included into
calculations to get an adequate description of the
electrodisintegration. The first relativistic model based on a separable kernel approach was Graz II \cite{Rupp:1989sg}. However, it
was impossible in principle to calculate FSI using it \cite{Bondarenko:2008mm}.
To solve this problem new relativistic
separable kernels \cite{Bondarenko:2010qv,Bondarenko:2008mm} were
elaborated. We apply them to the deuteron electrodisintegration including FSI in this paper.

The outgoing nucleons are described by the BS amplitude which can
be written as a sum of two terms:
\begin{eqnarray}
&&\psi_{SM_S}(p,p^*;P) = \psi^{(0)}_{SM_S}(p,p^*;P) \label{psi_all}\\
&&+\frac{i}{4\pi^3}
S_2(p;P)\int d^4k ~V(p,k;P)\psi_{SM_S}(k,p^*;P).\nonumber
\end{eqnarray}
The first term
\begin{eqnarray}
\psi_{SM_S}^{(0)} (p,p^*;P)=
(2\pi)^4\chi_{SM_S}(p;P)\delta(p-p^*)\label{wf_PWA}
\end{eqnarray}
is related to the outgoing pair of free nucleons (PWA),
$\chi_{SM_S}$ is a spinor function for two fermions. The second term in
(\ref{psi_all}) corresponds to the final state interaction of the
outgoing nucleons. It can be expressed through the $T$ matrix if
we use the following relation:
\begin{eqnarray}
&&\int d^4k~V(p,k;P)\psi_{SM_S}(k,p^*;P)=\\
&&\int d^4k~T(p,k;P)
\psi^{(0)}_{SM_S}(k,p^*;P)\nonumber
\end{eqnarray}
and can be rewritten as
\begin{eqnarray}
&&\psi_{SM_S}^{(t)}(p,p^*;P)= \label{psi_t}\\
&& \hspace*{10mm} 4\pi i S_2(p;P)
T(p,p^*;P)\chi_{SM_S}(p^*;P),\nonumber
\end{eqnarray}
here $(t)$ means that this part of the $np$ pair wave function is
related to the $T$ matrix. Applying the partial-wave decomposition
of the $T$ matrix (\ref{bse_spd}) the expression (\ref{psi_t}) can
be written as follows:
\begin{eqnarray}
&&\psi_{SM_S}^{(t)}(p,p^*;P)=  4\pi i \times \label{psi_t_decomposed}\\
&&\sum_{LmJMa} C_{LmSM_S}^{JM} Y_{Lm}^*(\hat\pp^*){\cal
Y}_{aM}(\pp) \phi_{a,J:LS+}( p_0,|\pp|; s), \nonumber
\end{eqnarray}
where $p^*=(0,\pp^*)$ with $|\pp^*|=\sqrt{s/4-m^2}$ is the
relative momentum of on-mass-shell nucleons in CM, $\hat\pp^*$
denotes the azimuthal angle $\thetaps$ between the $\pp^*$ and
$\q$ vectors and zenithal angle $\phi$. Since only
positive-energy partial-wave states are considered here the radial part
is:
\begin{eqnarray}
\phi_{a,J:LS+}(p_0,|\pp|; s)=\frac
{t_{a,J:LS+}(p_0,|\pp|;0,|\pp^*|; s)}{(\sqrt
s/2-\ep+i\epsilon)^2-p_0^2}.
\end{eqnarray}
According to definition (\ref{an_pa}) spin-angle functions ${\cal
Y}$ can be written as a product of Dirac $\gamma$ matrices in the
matrix representation \cite{Bondarenko:2002zz} as:
\begin{eqnarray}
&&{\cal Y}_{aM}(\pp)= \label{sap_np}\\
&&\frac{1}{\sqrt{8\pi}}
\frac{1}{4\ep(\ep+m)}(m+{\slashed p}_1)(1+\gamma_0){\cal G}_{aM}
(m-{\slashed p}_2),\nonumber
\end{eqnarray}
matrices ${\cal G}_{aM}$ are given in Table
\ref{tab:1}.
Decomposition (\ref{psi_t_decomposed})
is considered in detail in \cite{Bondarenko:2004pn}.
\begin{table}[h]
\begin{center}
{\renewcommand{\tabcolsep}{0.5mm}
\begin{tabular}{cc}
\hline\hline
$a={\footnotesize \left\{^{2S+1}L_J^\rho\right\}^{{\phantom{1}}^{\phantom{1}}} }$ & ${\cal G}_{aM}$ \\
\hline
$~{^1S_0^+}^{{\phantom{1}}^{\phantom{1}}}$ & $-\gamma_5$ \\
$~^3S_1^+$ & ${\slashed \xi}_{M}$ \\
$~^1P_1^+$ & $\frac{\sqrt{3}}{|\pp|}(p_1\cdot\xi_{M})\gamma_5$ \\
$~^3P_0^+$ & $-\frac{1}{2|\pp|}({\slashed p_1}-{\slashed p_2})$ \\
$~^3P_1^+$ &
$-\sqrt{\frac{3}{2}}\frac{1}{|\pp|}\left[(p_1\cdot\xi_{M})-\frac{1}{2}
{\slashed \xi}_{M}({\slashed p_1}-{\slashed p_2})\right]\gamma_5$ \\
$~^3D_1^+$ & $\frac{1}{\sqrt{2}}\left[{\slashed
\xi}_{M}+\frac{3}{2}\frac{1}{\pp^2}(p_1\cdot\xi_{M})({\slashed p}_1-{\slashed
p}_2)\right]$
\\
\hline\hline
\end{tabular}
}
\caption{Spin-angular parts ${\cal G}_{aM}$ (\ref{sap_np}) for the
$np$ pair; $p_1=(\ep,\pp)$, $p_2=(\ep,-\pp)$ are on-mass-shell
momenta, $\ep=\sqrt{\pp^2+m^2}$; $\gamma$ matrices are defined as in \cite{Bjorken}. }\label{tab:1}
\end{center}
\end{table}
Using definition (\ref{psi_all}) and substituting (\ref{wf_PWA}),
(\ref{psi_t_decomposed}) into (\ref{cur_fsi}), the final
expression for hadron current $<np:SM_S|j_\mu|d:1M>$ can be
obtained. It consists of two parts. One of them:
\begin{eqnarray}
&&<np:SM_S|j_\mu|d:1M>^{(0)}= \label{cur_0}\\
&&i\sum_{n=1,2}\left\{\Lambda({\cal
L}^{-1})\bar\chi_{SM_S}\left({p^*}\cm; P\cm\right)\Lambda({\cal
L})\Gamma_\mu^{(n)}(q) \right.\times \nonumber\\
&&S^{(n)}\left(\frac{K_{(0)}}{2}-(-1)^n
p^*-\frac{q}{2}\right)\left.\Gamma^{M}\left(p^*+(-1)^n\frac{q}{2};
K_{(0)}\right)\right\}\nonumber
\end{eqnarray}
corresponds to the electrodisintegration in PWA. Another one:
\begin{eqnarray}
&&<np:SM_S|j_\mu|d:1M>^{(t)}= \label{cur_t} \\ &&\frac{i}{4\pi^3}\sum_{n=1,2}\sum_{LmJM_JL'lm'}
C_{LmSM_S}^{JM_J} Y_{Lm}(\hat\pp^*)\times\nonumber\\
&&\int\limits_{-\infty}^\infty dp_0\cm \int\limits_0^\infty
{(\pp\cm)}^2 d|\pp\cm| \int\limits_{-1}^1
d\cos\thetap\cm \int\limits_0^{2\pi}d\phi\times \nonumber\\
&&{\rm Sp}\left\{\Lambda({\cal L }^{-1})\bar{\cal
Y}_{JL'SM_J}(\pp\cm) \Lambda({\cal L })
\Gamma_\mu^{(n)}(q)\times \right.\nonumber \\
&& \left. S^{(n)}\left(\frac{K_{(0)}}{2}-(-1)^np-\frac
q2\right){\cal Y}_{1lSm'}\left(\pp+(-1)^n\frac{\q}{2}\right)\right\}\times \nonumber\\
&& \frac {t_{L'L}^*(p\cm_0,|\pp\cm|;0,|\pp^*|; s)}{(\sqrt
s/2-\ep+i\epsilon)^2-p_0^2} \times \nonumber\\
&& g_l
\left(p_0+(-1)^n\frac{\omega}{2},\pp+(-1)^n\frac{\q}{2};K_{(0)}\right)
\nonumber
\end{eqnarray}
corresponds to the process when FSI is taken into account. Here
$g_l$ is a radial part of the deuteron vertex function $\Gamma^M$. The part
${\rm Sp}\left\{\ldots\right\}$ has been calculated using the
algebra manipulation package MAPLE. The three-dimensional integration over $p_0\cm$, $|\pp\cm|$ and
$\cos\thetap\cm$ has been performed numerically using the
programming language FORTRAN.
\begin{table}
\begin{center}
{\renewcommand{\tabcolsep}{1mm}
\begin{tabular}{|ll|c|c|c|}
\hline
            &         &  set I \cite{Bussiere:1981mv}   &  set II \cite{Bussiere:1981mv}   &  \cite{TurckChieze:1984fy}     \\
\hline $E_e$, GeV
            &         &  0.500  &  0.500  &  0.560    \\
\hline $E_e^\prime$, GeV
            &         &  0.395  &  0.352  &  0.360   \\
\hline
$\theta$, ${}^{\circ}$
            &         &   59    &  44.4   &    25    \\
\hline $\pp_n$, GeV/$c$
            & min &  0.005  &  0.165  &  0.294   \\
            & max &  0.350  &  0.350  &  0.550   \\
\hline
$\theta_n$, ${}^{\circ}$
            & min & 101.81  &  172.07 & 153.01   \\
            & max &  37.78  &  70.23  &  20.81   \\
\hline
$\theta_{qe}$, ${}^{\circ}$
            & min &  48.79  &  44.74  &  33.06   \\
\hline $\pp_p$, GeV/$c$
            & min &  0.451  &  0.514  &  0.557   \\
            & max &  0.276  &  0.403  &  0.306   \\
\hline
$\theta_{p}$, ${}^{\circ}$
            & min &  0.622  &   2.54  &  13.86   \\
            & max &  51.03  &  54.90  & 140.28   \\
\hline
$\theta_{pe}$, ${}^{\circ}$
            & min &  49.41  &  47.28  &  46.92   \\
            & max &  99.81  &  99.64  & 173.35   \\
\hline $\sqrt s$, GeV
            &     &  1.929  &  1.993  &  2.057    \\
\hline $\sqrt s- 2m$, GeV
            &     &  0.051  &  0.115  &  0.176   \\
\hline $Q^2$, (GeV/$c$)$^2$
            &     &  0.192  &  0.101  &  0.038   \\
\hline $\omega$, GeV
            &     &  0.105  &  0.148  &  0.200   \\
\hline $|\q|$, GeV/$c$
            &     &  0.450  &  0.350  &  0.279   \\
\hline
\end{tabular}
} \caption{ {\small Kinematic conditions considered in the paper.
Here all quantities are in LS. In addition to those which are
defined in the text, they are: angle $\theta_{qe}$ between the
beam and the virtual photon; neutron momentum $\pp_n$ and angle
$\theta_n$ between the neutron and the virtual photon
($\pp_p,\theta_p$ -- the same for the proton); $\theta_{pe}$
($\theta_{qe}$) -- the angle between the beam and the proton
(virtual photon).} }\label{tab:2}
\end{center}
\end{table}
\begin{figure}[h]
\includegraphics[width=1\linewidth]{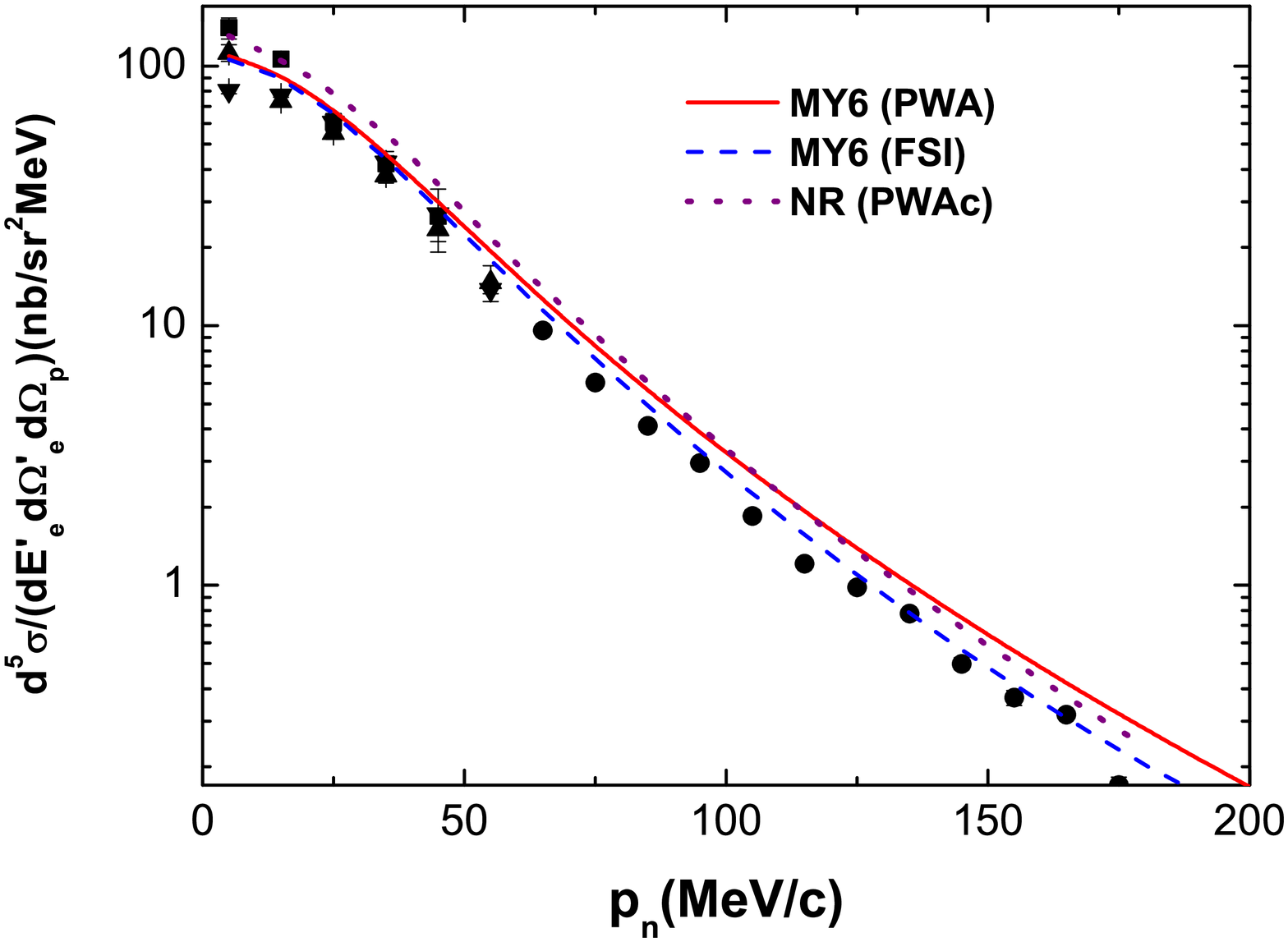}
\caption{Cross section (\ref{3cross_0}) depending on recoil
neutron momentum $|\pp_n|$ calculated under kinematic conditions set I
of the Sacle experiment \cite{Bussiere:1981mv}. The
notations are following: MY6 (PWA) (red solid line) -
relativistic calculation in the plane-wave approximation with the
MY6 potential \cite{Bondarenko:2010qv}; MY6 (FSI) (blue dashed line)
- relativistic calculation including FSI effects; NR (PWAc)
(violet dotted line) - nonrelativistic calculation
\cite{Shebeko}.}\label{fig:1}
\end{figure}
\begin{figure}[h]
\includegraphics[width=1\linewidth]{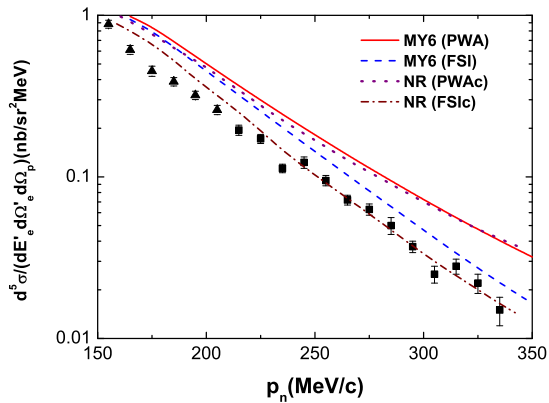}
\caption{The same as in Fig.\ref{fig:1} but under kinematic
conditions set
II of the Sacle experiment \cite{Bussiere:1981mv}. The nonrelativistic calculation NR (FSIc) (brown
dashed-dotted line) includes FSI effects.}\label{fig:2}
\end{figure}
\begin{figure}[h]
\includegraphics[width=1\linewidth]{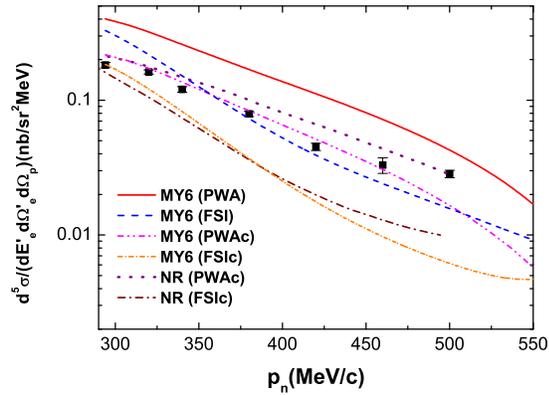}
\caption{The same as in Figs.\ref{fig:1},\ref{fig:2} but under
kinematic conditions of the Sacle experiment
\cite{TurckChieze:1984fy}. Two additional results are presented
for comparison: MY6 (PWAc) (pink dashed-dotted-dotted line) -
relativistic PWA calculation; MY6 (FSIc) (orange dashed-dotted line) -
relativistic calculation with FSI effects; both obtained under current
conservation condition (\ref{curcon}).}\label{fig:3}
\end{figure}
\section{Results and discussion}\label{sec:5}
The differential cross section (\ref{3cross_0}) is calculated under
three kinematic conditions of the Sacle experiment
\cite{Bussiere:1981mv,TurckChieze:1984fy} (described in Table \ref{tab:2}) and is present in
Figs.\ref{fig:1}-\ref{fig:3}. The calculations have been performed
within the relativistic impulse approximation for two different
cases: when the outgoing nucleons are supposed to be free (PWA)
and when the final state interaction between the nucleons is taken
into account (FSI). The partial-wave states of the $np$ pair with
total angular momentum $J=0,1$ have been considered. The used
relativistic model consists of two parts: the separable potential MY6
\cite{Bondarenko:2010qv} for the bound (deuteron) and scattered
$^3S_1$-$^3D_1$ states and separable potentials of various ranks
\cite{Bondarenko:2008mm} - for all the other partial-wave states
($^1S_0$, $^1P_1$, $^3P_0$, $^3P_1$). The obtained results have been
compared with the nonrelativistic model \cite{Shebeko}.

In Fig.\ref{fig:1}, relativistic MY6 (PWA), MY6 (FSI) and
nonrelativistic NR (PWAc) calculations under kinematic
conditions \cite{Bussiere:1981mv} (set I) are practically coincide
with each other and go close to the experimental data. The
notation "c" in NR (PWAc) means that the corresponding hadron
current satisfies the current conservation condition
\cite{Forest:1983vc}:
\begin{eqnarray}
q^\mu J_\mu=0. \label{curcon}
\end{eqnarray}
So, it can
be concluded that relativistic and FSI effects do not play an
important role in description of the deuteron electrodisintegration
at low energies.

Fig.\ref{fig:2} considers the cross section (\ref{3cross_0})
calculated under kinematic conditions \cite{Bussiere:1981mv} (set
II). Here FSI effects decrease the resulting cross section
noticeably. A slight difference is seen between nonrelativistic
and relativistic results.

The cross section under kinematic conditions
\cite{TurckChieze:1984fy} is presented in Fig.\ref{fig:3}. In this
case PWA and FSI calculations differ significantly. It means that
the influence of FSI on observables increases with increasing
$|\pp_n|$. To investigate relativistic effects, two additional
calculations MY6 (PWAc), MY6 (FSIc) have been performed when the
condition (\ref{curcon}) has been taken into account. The
relativistic one-particle current within the impulse approximation
(\ref{cur_fsi}) does not satisfy (\ref{curcon}). We impose this
condition on the hadron current (\ref{cur_0}), (\ref{cur_t}) as it
was done in \cite{Shebeko} for a nonrelativistic model to compare our relativistic result with
it and investigate relativistic effects. It has been done only under
kinematic conditions \cite{TurckChieze:1984fy} since an influence
of the current conservation condition becomes significant at high
outgoing neutron momenta $|\pp_n|$. The comparison of MY6 (PWAc)
and NR (PWAc), MY6 (FSIc) and NR (FSIc) demonstrates that relativistic
effects become significant with increasing $|\pp_n|$. FSI effects
decrease the obtained cross section.

As it is seen from the figures, an influence of FSI increases
with increasing outgoing neutron momentum $|\pp_n|$. FSI effects
are significant at outgoing neutron momenta $|\pp_n|>150$\,MeV/c
even at low momentum transfer. It has been demonstrated that the
FSI contribution always decreases the value of the cross section
(\ref{3cross_0}). From present results it can also be seen that
the FSI contribution relatively to the PWA one is approximately
the same both for nonrelativistic and relativistic
models. However, other effects, like negative-energy partial-wave
states and two-body currents, should be taken into account for
further conclusions. It should be also emphasized that the
relativistic calculation of FSI has been performed
within the BS approach using a relativistic separable potential for the first time.
Recently the considered separable potentials
\cite{Bondarenko:2010qv,Bondarenko:2008mm} have been modified to
take into account inelastic processes (production of various
mesons) occurring in the NN interaction at high energies
\cite{Bondarenko:2011hs}. However, energies of nucleons do not
reach the inelasticity threshold under kinematic conditions
\cite{Bussiere:1981mv}. The inelasticity effects are nonzero under
conditions \cite{TurckChieze:1984fy} but they are conjectured to
be small in this case.

\end{document}